\PassOptionsToPackage{bookmarks=false}{hyperref}
\documentclass[sigconf]{acmart}

\AtBeginDocument{%
  \providecommand\BibTeX{{%
    \normalfont B\kern-0.5em{\scshape i\kern-0.25em b}\kern-0.8em\TeX}}}

\copyrightyear{2020} 
\acmYear{2020} 
\setcopyright{acmlicensed}
\acmConference[ICSE-NIER'20]{New Ideas and Emerging Results}{May 23--29, 2020}{Seoul, Republic of Korea}
\acmPrice{15.00}
\acmDOI{10.1145/3377816.3381744}
\acmISBN{978-1-4503-7126-1/20/05}

\graphicspath{{figures/}}
\DeclareGraphicsExtensions{.eps,.pdf,.png}

\usepackage{subcaption}

\usepackage[export]{adjustbox} 
\usepackage{amssymb,amsmath}

\usepackage{hyperref}

\usepackage{enumitem}

\usepackage{calc}

\usepackage[export]{adjustbox}

\usepackage[framemethod=tikz]{mdframed}
\newmdenv[
  innerleftmargin=7pt,
  innerrightmargin=7pt,
  tikzsetting={draw=black,dashed,line width=0.5pt,dash pattern = on 4pt off 2pt},
  linecolor=white,
  backgroundcolor=white
]{dashedbox}
\newmdenv[
  innerleftmargin=7pt,
  innerrightmargin=7pt,
  tikzsetting={draw=black, line width=0.5pt},
  linecolor=black,
  backgroundcolor=white
]{normalbox}
\newmdenv[
  topline=false,
  bottomline=false,
  rightline=false,
  skipabove=\topsep,
  skipbelow=\topsep,
  innertopmargin=0pt,
  innerbottommargin=0pt,
  innerleftmargin=7pt,
  innerrightmargin=0pt,
  tikzsetting={draw=black, line width=3pt},
  linecolor=black,
  backgroundcolor=white
]{verticalline}

\usepackage{listings}

\definecolor{light-gray}{gray}{0.97}
\definecolor{gray}{rgb}{0.4,0.4,0.4}
\definecolor{darkblue}{rgb}{0.0,0.0,0.6}
\definecolor{cyan}{rgb}{0.0,0.6,0.6}

\usepackage{tabularx}

\usepackage{multirow}

\definecolor{VeryLightGray}{rgb}{0.92,0.92,0.92}
\usepackage{colortbl}

\usepackage{bm} 

\usepackage{nicefrac}

\usepackage{amsmath}
\usepackage{amssymb} 
\usepackage{eucal} 

\usepackage{siunitx}
\begin{document}

\title{Code Duplication on Stack Overflow}

\author{Sebastian Baltes}
\orcid{0000-0002-2442-7522}
\email{sebastian.baltes@adelaide.edu.au	} 
\affiliation{%
  \institution{The University of Adelaide, Australia}
}

\author{Christoph Treude}
\orcid{0000-0002-6919-2149}
\email{christoph.treude@adelaide.edu.au}
\affiliation{%
  \institution{The University of Adelaide, Australia}
}


\begin{abstract}
Despite the unarguable importance of Stack Overflow (SO) for the daily work of many software developers and despite existing knowledge about the impact of code duplication on software maintainability, the prevalence and implications of code clones on SO have not yet received the attention they deserve.
In this paper, we motivate why studies on code duplication within SO are needed and how existing studies on code reuse differ from this new research direction.
We present similarities and differences between code clones in general and code clones on SO and point to open questions that need to be addressed to be able to make data-informed decisions about how to properly handle clones on this important platform.
We present results from a first preliminary investigation, indicating that clones on SO are common and diverse.
We further point to specific challenges, including incentives for users to clone successful answers and difficulties with bulk edits on the platform, and conclude with possible directions for future work.
\end{abstract}



\maketitle

\section{Introduction}

Code clones have been extensively studied in the software engineering research community.
Juergens et al. found that inconsistent code clones can be a major problem during the development and maintenance of software projects unless \emph{``special care is taken to find and track existing clones and their evolution''}~\cite{juergens_code_2009}.
Stack Overflow (SO) threads, often containing code snippets together with explanations~\cite{yang_query_2016}, serve as an important crowd-sourced software documentation resource~\cite{parnin_crowd_2012, treude_how_2011}.
Despite the fact that code clones on SO can suffer from similar issues as code clones in software projects, it is only recently that researchers started investigating them.
Studies have shown that developers utilise code snippets from SO in their software projects, regardless of maintainability, security, and licensing implications~\cite{baltes_attribution_2017, baltes_usage_2019, an_stack_2017, yang_stack_2017,  gharehyazie_here_2017, abdalkareem_code_2017, xia_what_2017, fischer_stack_2017, acar_you_2016, wu_how_2018}.
The main focus of that previous work was, however, to study how and why developers (re-)use SO code snippets outside of the question-and-answer platform.
While researchers worked on identifying duplicate questions~\cite{zhang_multi-factor_2015, ahasanuzzaman_mining_2016, zhang_detecting_2017}, their main goal was to replace or support the manual moderator process for marking duplicate questions rather than supporting the maintenance and evolution of code on SO.
Considering the importance that SO has today for the daily work of many software developers worldwide and the fact that in many posts, non-trivial code snippets are collected and maintained, it is surprising that SO does not have proper features for code versioning and bug tracking.
Text and code are versioned together as Markdown content~\cite{baltes_sotorrent:_2018-1}, making it hard to identify changes to the code snippets in the provided revision view. 
Furthermore, there is no language-specific syntax highlighting or error checking in SO's online Markdown editor, leading to many snippets that are not parseable, compilable, or even runnable~\cite{yang_query_2016}.
Finally, there is no way to report bugs in SO code snippets other than posting a comment or an alternative answer.
Despite the above-mentioned challenges, code \emph{is} maintained and \emph{does} evolve on SO~\cite{baltes_sotorrent:_2018-1}.

The purpose of this article is to point the research community to open questions regarding code clones on SO and to motivate how research in that area could inform significant improvements for the platform.
We present a preliminary analysis of code clones within SO and point to directions for future work.
As outlined above, the external usage of code snippets from SO has been studied in-depth.
Therefore, our main focus is on code clones \emph{within} the platform (see Section~\ref{sec:research-questions} for our specific research questions).
While we are in an early stage of this research project, we already shared preliminary results with the SO community and started a discussion about how to handle code clones on the platform.
The corresponding thread\footnote{\url{https://meta.stackoverflow.com/q/375761}} attracted over 1,000 views and got upvoted to a score of 45.
We will use the community's feedback to guide our next steps.
Our vision is that a thorough study of code clones on SO together with the ongoing discussion in the SO community will lead to revised recommendations for authors and improved tool support for handling those clones.


\section{Motivating Examples}
\label{sec:examples}

In this paper, we want to discuss two use cases of duplicated code on SO that have not been adequately targeted by previous research.
The first case covers content that was originally posted on SO, without any indication of an external source.
For such snippets, the (visible) evolution mainly takes place on the SO platform, but nevertheless maintenance issues may arise.
The second case is content that was copied from external sources into SO posts.
Such clones suffer from the same maintenance issues as the clones mentioned above, but the additional external availability makes the evolution even more complex and may further introduce licensing issues.


\begin{figure}
\centering
\includegraphics[width=1\columnwidth,  trim=0.0in 0.0in 0.0in 0.0in]{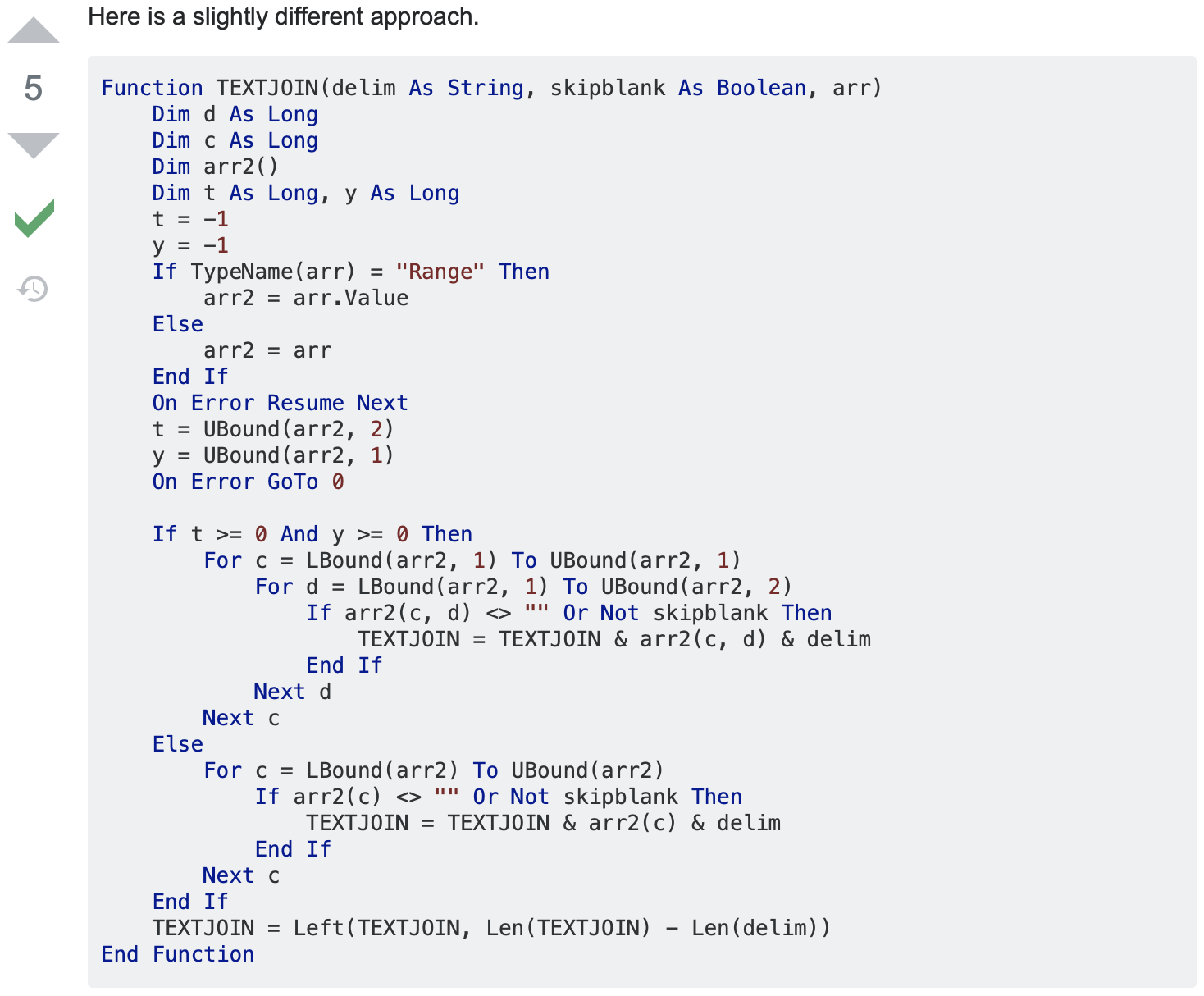} 
\caption{First usage of a VBA code snippet on 16 September 2016 in answer \href{https://stackoverflow.com/a/39532855}{39532855},
the most recent usage by the same user was on 27 August 2018 in answer \href{https://stackoverflow.com/a/52040136}{52040136}.
Within that timeframe, 31 copies of the same snippet were posted on SO, most of them by the same user. Consistently changing the snippet would require a manual update of all those copies.}
\label{fig:case-1}
\end{figure}

\subsection{Case 1: Original code snippets being reused in multiple threads}
\label{sec:case-1}

Figure~\ref{fig:case-1} shows the first usage of a VBA (Visual Basic for Applications) code snippet, of which, within a time span of two years, 31 exact copies were posted on SO.
Except for three questions, the snippet was exclusively used in different answers posted by the same user.
None of those copies reference each other, meaning that an update in one copy would stay unnoticed in the other threads.
The original author of the snippet managed to attract an accumulated number of over 25,000 views and an accumulated score of 65.
Interestingly, it is not the initial answer that attracted the most views and the highest score.
When we proposed to link more recent posts to the first occurrence, the author rejected our edit with a generic reply.\footnote{\url{https://stackoverflow.com/review/suggested-edits/21495979}}
This leads to our first observation related to code duplication on SO:

\vspace{0.5\baselineskip}
\begin{verticalline}
SO users may utilise code clones to accumulate views and upvotes. At the same time, they can reject edit proposals referencing the clones.
\end{verticalline}

Note that we referred to exact copies in this example.
It is likely that there exist further type-2 clones (e.g., with renamed identifiers) or type-3 clones (e.g., with added or removed statements) of the snippet. 
Some of those clones may contain fixes not yet propagated to the other clones.

\subsection{Case 2: Externally available snippets being reused in multiple threads}
\label{sec:case-2}

Not all snippets on SO are originally posted there, many are copied to and from the platform~\cite{baltes_usage_2019, xia_what_2017, wu_how_2018}.
While this observation is not new, many existing studies focused on the implications of reusing content from SO, but not so much on the licensing and maintainability implications for the platform itself.

Considering SO's role in the software documentation landscape, it is not surprising that content from reference documentation resources is being reused on SO.
Even in cases where a license-compatible usage of content would be straightforward, SO authors fail to adhere to license requirements.
For example, a Java snippet about server certificate verification from an official Android tutorial\footnote{\url{https://developer.android.com/training/articles/security-ssl.html}} has been copied (at least) 14 times into SO. 
This happened in a timespan of over four years.
Google licensed this tutorial---and thus the snippet---under CC BY 2.5.
This license is compatible with SO's CC BY-SA license but requires attribution, which SO authors do not always provide.
Besides this licensing aspect, in the above example, there is unarguably one authoritative source for the snippet where bug fixes or updates would be posted.
Therefore, our next observation is:

\vspace{0.5\baselineskip}
\begin{verticalline}
SO users copy code snippets from reference documentation into SO posts.
Besides licensing and copyright implications, it is questionable whether this behaviour contributes to the sustainability of the platform, because users may reuse outdated information in case the authoritative source gets modified.
Even if one of the clones gets updated, the changes are not automatically propagated within the network of clones on SO.
\end{verticalline}

There is another angle to this case: We added the missing attribution in two posts containing the above-mentioned snippet from the official Android documentation.
While those edits were accepted, there is a rate limited for such bulk edits.\footnote{\url{https://meta.stackexchange.com/a/281202}}
Hence, we observe that:

\vspace{0.5\baselineskip}
\begin{verticalline}
Due to SO's rate limiting, missing attribution cannot be easily added to posts.
Tools that researchers build can currently only very slowly propagate proposed changes to affected posts. 
\end{verticalline}

\section{Research Questions}
\label{sec:research-questions}

Motivated by the two cases outlined above, we argue that the following research questions need to be addressed to be able to make an informed decision on how to handle code clones on SO.
The goal of this research is to provide data-informed actionable recommendations which the community and SO's internal team can use to update their user guides, but also to build tool support for managing and maintaining code clones on the platform.

\begin{itemize}[labelindent=\parindent, labelwidth=\widthof{\textbf{RQ1:}}, label=\textbf{RQ1:}, leftmargin=*, align=parleft, parsep=0pt, partopsep=0pt, topsep=1ex, noitemsep]
\item[\textbf{RQ1:}] How frequently are code snippets copied between SO posts?
\item[\textbf{RQ2:}] What types of clones exist and how are they related?
\item[\textbf{RQ3:}] What are typical external sources of code snippets on SO and which licensing and maintainability implications are associated with those sources?
\end{itemize}

The first research question helps us to assess how common the outlined problem is, the second to deepen our understanding of the different use cases of clones on SO, and the last to understand the implications beyond the platform.


\begin{figure*}
\includegraphics[width=0.35\textwidth,  trim=0.0in 0.4in 0.3in 0.2in]{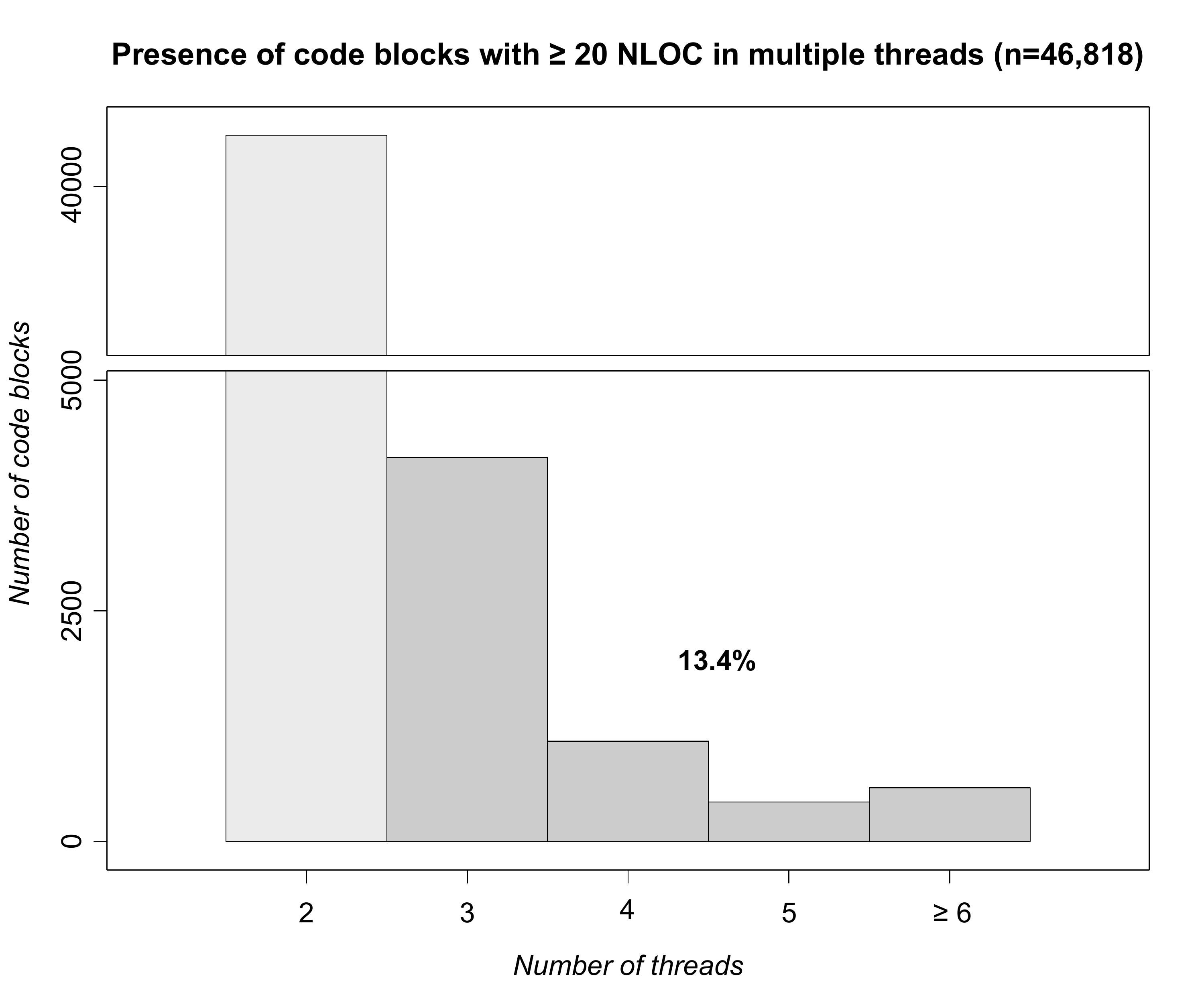} 
\hfill
\includegraphics[width=0.62\textwidth,  trim=0.0in 0.5in 0.0in 0.0in]{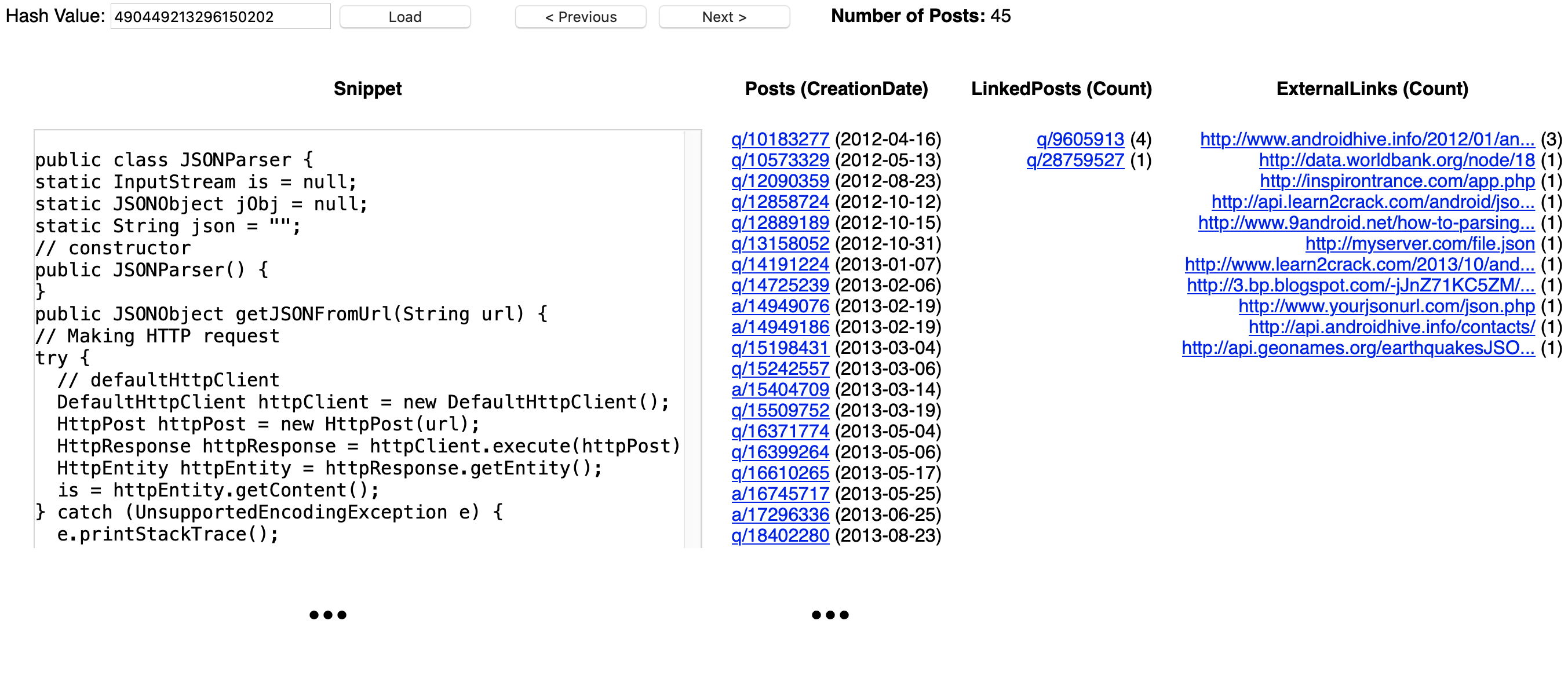} 
\caption{\emph{Left:} Copies of non-trivial code blocks ($\ge 20$ NLOC) in multiple SO threads. \emph{Right:} \href{http://research.sbaltes.com/so-clones/snippet-view.html?hashValue=490449213296150202}{Snippet view} of \texttt{so-clones} tool showing a code snippet that has likely been copied from the website \href{https://www.androidhive.info/2012/05/how-to-connect-android-with-php-mysql/}{\emph{androidhive}} into SO.}
\label{fig:code-clones-thread-count}
\label{fig:so-clones-snippet-view}
\end{figure*}

\section{Preliminary Analysis}

We conducted a preliminary analysis focusing on exact code clones on SO.
While this approach has the advantage of not being limited to a certain programming language, extending the analysis to also cover other types of clones will be a logical next step.

\subsection{Data Retrieval}

To detect code clones on SO, we utilised the BigQuery version of \emph{SOTorrent}~\cite{baltes_sotorrent:_2018-1}. 
First, we selected all code blocks from the most recent post versions and normalised the contained whitespace characters.
To this end, we: (1) replaced sequences of new lines with a single new line character, (2) removed new lines at the end of the last line, and (3) removed lines only containing brackets (\texttt{()[]\{\}}).
Using this normalised content, we calculated the normalised line count of those code blocks (NLOC).
To derive fingerprints of the snippets, we only considered alphanumeric characters 
and applied BigQuery's \texttt{FARM\_FINGERPRINT} function to the normalised code block contents.
This yielded 43,942,960 distinct fingerprints (i.e., normalised code blocks).
We then used this fingerprint to determine the posts which contain a certain snippet, aggregating that information per thread. 
To select cloned code blocks, we first selected the ones present in at least two different threads, yielding 909,323 distinct fingerprints.
This provides a first estimate for \textbf{RQ1}: 2.1\% of all distinct code blocks have a copy in another thread.
To select only non-trivial code snippets, we first used a threshold of six normalised lines of code, as proposed by Bellon et al.~\cite{bellon_comparison_2007}.
We ranked the remaining 215,746 code snippets according to the number of threads they were found in and according to their normalised length.
Then, we qualitatively analysed the first 50 snippets in that list.
Since we considered 25 of those snippets to be either too trivial or non-code, we decided to adjust the threshold for the normalised line count to 20.
The stricter filtering led to a second sample with 46,818 code snippets.
Those snippets had an average length of 42.6 normalized lines ($\textit{SD}=37.7$, $\textit{Mdn}=30$, $\textit{IQR}=22$) and were present in 2.3 different threads ($\textit{SD}=1.1$, $\textit{Mdn}=2$, $\textit{IQR}=0$); 13.4\% of the snippets were present in more than two threads (see Figure~\ref{fig:code-clones-thread-count}).
We provide the retrieval scripts and the coding for both samples ($\ge 6$ NLOC and $\ge 20$ NLOC) on Zenodo.\footnote{\url{https://doi.org/10.5281/zenodo.1474222} and \url{https://doi.org/10.5281/zenodo.3596367}}

\subsection{Qualitative Analysis}

To address \textbf{RQ2} and \textbf{RQ3}, we first ranked the code snippets according to their thread count and length and then qualitatively analysed the first 50 snippets according to that ranking using a web tool\footnote{\url{https://doi.org/10.5281/zenodo.1474207}} we specifically designed for that purpose. 
The tool allows to focus on a single snippet in a dedicated view, showing the snippet, its fingerprint, the posts containing the snippet sorted by their creation date, other posts linked from those posts, and linked external sources (see Figure~\ref{fig:so-clones-snippet-view}).
The latter information helped us to identify if and from where a snippet may have been copied into SO.
While we still categorised ten snippets as configuration files, 29 snippets were non-trivial source code snippets (mainly Java and VB/VBA).
Other categories included XML GUI definitions for Android, JSON/XML examples, and HTML files.
Except for two cases, we were able to identify the (or at least a) source of the snippet by following links in the posts and searching for parts of the snippets online.
Only in four cases, we considered the snippets to be originally from SO.
The main external sources were a website providing Android tutorials (\href{https://www.androidhive.info/}{\emph{androidhive.info}}, ten snippets) and the official Android documentation (\href{http://developer.android.com/}{\emph{developer.android.com}}, four snippets).
We identified possible licensing conflicts in 31 cases, either because the website did not provide a license or because the content was distributed under a restrictive license or restrictive terms of use.
In the following, we describe the two main external sources in more detail.

The independent Android website \emph{androidhive} has rather restrictive terms of use. 
Nevertheless, only few posts attribute this source (3 out of 45 posts in the example shown in Figure~\ref{fig:so-clones-snippet-view}).
It is unclear whether the snippet has actually been copied from this external source since the creation of the posts on \emph{androidhive} and SO were both around April/May 2012.
If the 45 snippets were copied into SO, their usage would be problematic.
In fact, we identified four more variants of that same code snippet among the 50 snippets we analysed.
If SO is the original source, the usage on \emph{androidhive} does not adhere to SO's CC BY-SA license~\cite{baltes_usage_2019}.
The snippets copied from the \emph{official Android documentation} are licensed under CC BY 2.5. 
This license allows usage under SO's CC BY-SA license, but only when attributing the original source.
However, users often did not add a link to the Android documentation to their posts.
Thus, also those usages could lead to \emph{licensing issues}.

Leaving the licensing implications aside, code clones within SO are also problematic for the platform's \emph{maintainability} and \emph{usability}.
Code duplicates could, for example, indicate that different threads solved a similar problem.
If there is no link between the threads, information is spread over the platform and hard to capture for readers.
Another example is SO's recommendation to \emph{``always quote the most relevant part of an important link, in case the target site is unreachable or goes permanently offline''}.\footnote{\url{https://stackoverflow.com/help/how-to-answer}}
While it makes sense to quote important aspects of external sources, it can be questioned whether it is reasonable to maintain several independent copies of external code snippets on SO.
Assuming that a snippet in the reference documentation is updated, all copies on SO would require a manual update as well. 

\subsection{Community Involvement}

To involve the community and discuss how to best approach those licensing, maintainability, and usability issues, we created a post on SO Meta.
We outlined the two cases presented in Sections~\ref{sec:case-1} and \ref{sec:case-2} using examples from our preliminary analysis and asked the community: \emph{How to handle code clones on Stack Overflow?}

One preliminary observation from the discussion is that the community seems to be in favour of adding missing attribution to SO posts where it is missing.
This would enable tool support for automatically checking external sources for updated versions of snippets, but only solve the licensing issue for snippets licensed under a rather permissive license.
Regarding approaches to handle clones within SO, there is no clear opinion yet.
Besides continuing to work on the research questions presented in Section~\ref{sec:research-questions}, we will update the discussion and finally implement the approach that the community prefers.
Possible directions include to automatically propose post edits adding missing attribution, to automatically link to the first occurrence in a set of duplicates, or to mark threads as related based on the similarity of the code blocks they contain.
The existing \href{https://github.com/SOBotics/Guttenberg}{Guttenberg bot} monitors newly posted questions and answers and automatically reports plagiarism, mainly based on String similarity between posts.
The bot does, however, compare whole posts without isolating code snippets first.
When we ran our study, the bot was already active.
Nevertheless, we still found a considerable amount of duplicated code on the platform. 
We have shared our results with the Guttenberg team. 

\section{Conclusion}




With this paper, we want to point to the fact that code clones on SO, similar to clones in regular software projects, affect the maintainability of posts and can lead to licensing issues.
However, we also point to differences such as the fact that SO users may be encouraged to clone successful answers to achieve a higher reputation and that snippets are difficult to modify through bulk edits.
These differences have practical implications and might suggest new features for SO, such as allowing bulk edits for adding attribution information or improving detection of overly zealous self-plagiarism.
We also mentioned that SO's current recommendation for handling content from external resources may not be suitable for code taken from reference documentation, because the authoritative source should be the reference documentation and not SO.
Moreover, when clones on SO are not updated, users consulting SO threads instead of the reference documentation may be prone to using outdated or erroneous information.

Our preliminary results suggest that code duplication on SO is relatively common (\textbf{RQ1}).
Despite our limitation on exact clones, we found that 2.1\% of all unique code blocks were used in more than one thread.
Moreover, when we focused on unique code blocks with $\ge 20$ NLOC, we still found 47k of them being used in more than one thread.
Our analysis also revealed that a wide variety of snippets is being cloned (\textbf{RQ2}).
We noticed cases where the same user copied the same snippet into several answers in different threads.
As motivated in Section~\ref{sec:examples}, future work should investigate this behaviour.
Further, the external sources of code snippets on SO (\textbf{RQ3}) range from random blog posts to official reference documentation pages (see our Android example).
Identifying typical external sources and developing ways to keep the content on SO up-to-date is a promising direction for future work.
This research could even be extended to include other online forums or code hosting platforms.


\bibliographystyle{IEEEtran}
\bibliography{literature}

\end{document}